\documentclass[twocolumn,secnumarabic,amssymb, nobibnotes, aps, prd,superscriptaddress]{revtex4-2}

\usepackage{graphicx}
\usepackage{color}
\usepackage{xcolor}
\usepackage{amsmath}
\usepackage{gensymb}
\usepackage{tabularx}
\usepackage{multirow}
\usepackage{ulem}
\usepackage[left,modulo]{lineno}
\usepackage{upgreek}

\begin{document}
\title{Emergence of a Hump in the Cubic Dielectric Response of Glycerol}

\author{Marceau H\'enot}
\email[Corresponding author: ]{marceau.henot@cea.fr}
\affiliation{SPEC, CEA, CNRS, Université Paris-Saclay, CEA Saclay Bat 772, 91191 Gif-sur-Yvette Cedex, France.}

\author{François Ladieu}
\affiliation{SPEC, CEA, CNRS, Université Paris-Saclay, CEA Saclay Bat 772, 91191 Gif-sur-Yvette Cedex, France.}

\date{\today}
\begin{abstract}
We report a direct determination of the cubic dielectric spectra of a realistic polar molecule, glycerol, from molecular dynamic (MD) simulations. From the liquid state to the mildly supercooled regime, we observed the emergence and growth of a hump in the cubic modulus, traditionally associated with collective effects in the dynamics. Its evolution follows that of dynamical correlations probed by the four-point susceptibility and that of the activation energy of the relaxation time. In particular, its appearance at high temperature coincides with the onset of super-activation. We show that, for this system, the shape of cubic spectra is only weakly affected by dipolar cross-correlations. The good agreement with experimental observations, despite the difference in temperature range, demonstrates the relevance of this approach to help get an insight into the intricate effects probed by non-linear dielectric spectroscopy.
\end{abstract}

\maketitle

The cooling of a supercooled liquid is associated with a dramatic slow-down of its structural relaxation time, ultimately leading to a virtually arrested out-of-equilibrium state called a glass~\cite{angell2000relaxation}. This phenomenon is accompanied by an increasingly spatially correlated dynamics, as demonstrated experimentally~\cite{vidal2000direct, reinsberg2001length, crauste2010evidence} and numerically~\cite{donati1998stringlike, 2011bethierDH, scalliet2022thirty}. However, its physical origin, and in particular the role of a growing static length scale upon cooling~\cite{berthier2019can, biroli2021amorphous}, is still an ongoing question~\cite{wyart2017does, pica2024local}.

The precise experimental characterization of the correlated dynamic is a challenge. One popular technique in the literature is non-linear dielectric spectroscopy which focuses on the response of a polar liquid to a large electric field. While applying a tiny ac field allows to probe its reorientational dynamics in a non-perturbative way, a large field, by displacing the equilibrium state~\cite{johari2013effects}, gives access to rich and original information regarding cross-correlation effects~\cite{singh2012watching, richert2017nonlinear, gabriel2021high}, the out-of-equilibrium response to the perturbation~\cite{richert2021structural, richert2023fast} or the collective nature of the relaxation~\cite{crauste2010evidence, brun2011nonlinear, bauer2013cooperativity, lhote2014control, casalini2015dynamic, gadige2017unifying, albert2016fifth}. The later effect was predicted~\cite{bouchaud2005nonlinear}, before it was ever measured~\cite{crauste2010evidence}, to be associated with the presence of a hump in the non-linear modulus, which grows upon cooling, a feature recently observed also in non-linear shear-mechanical experiments~\cite{moch2024dielectric}. In this framework, the height of the hump is proportional to the number of correlated molecules $N_\mathrm{corr}$. Nevertheless, the precise link between the collective dynamics and the cubic response is hard to establish and propositions were made to interpret the cubic hump without relying explicitly (or at all) on collective effects~\cite{diezemann2012nonlinear, diezemann2013higher,  kim2016dynamics, diezemann2018nonlinear, speck2021modeling}. To get some further input in this interpretation issue, the way to probe non-linearities were widely varied: either by measuring the cubic contribution at frequencies $f$ and $3f$ to a large ac field at $f$~\cite{crauste2010evidence,brun2011nonlinear, gadige2017unifying, richert2017nonlinear}, or by probing the quadratic change in the ac linear response induced by a large dc field $E_0$~\cite{lhote2014control, samanta2015dynamics, samanta2016electrorheological, young2016field, gadige2017unifying, young2017nonlinear, gabriel2021high}. This method has the advantage of not being subject to the heating effect induced by a large ac field. The associated complex cubic susceptibility $\chi_{2,1}$ can be defined as~\cite{lhote2014control}:
\begin{align}
\label{eq_defchi21}
    \chi_{2,1}(f) = \frac{1}{2\times 3}\frac{\partial^2 \chi_1(f)}{\partial E_\mathrm{0}^2}
\end{align}
 Applying a strong static electric field has two direct consequences: a decrease in static susceptibility, which, for a dilute gaz of dipole, would be simply due to the saturation effect, and a slowdown of the dynamics, called electro-rheological effect~\cite{lhote2014control, samanta2016electrorheological}. Both effect can \textit{a priori} be made complicated by dipolar cross-correlation effects. Molecular dynamics (MD) simulations have proven useful to disentangle such effects in the linear response~\cite{koperwas2022computational, alvarez2023debye, henotPCCP2023}. A few MD works have accessed the non-linear responses of liquids~\cite{english2003hydrogen, marracino2022rationale, matyushov2015nonlinear, samanta2022nonlinear, matyushov2023nonlinear, sauer2024linear, woodcox2023simulating}. Yet, no cubic spectra comparable to experiments in glass-forming liquids have been extracted.

In this letter, we report the first direct computation of the non-linear dielectric response $\chi_{2,1}(f)$ from a 3D MD simulation of glycerol, a system widely studied experimentally. We observed the emergence and growth upon cooling of the characteristic hump in the cubic modulus which compares well with experimental observations despite the difference in accessible temperature range. We take advantage of the numerical approach to study the effect of cross-correlations on the cubic response. Finally, we evidence a similar behavior for the cubic hump, a direct characterization of the dynamical heterogeneities, and a proxy of $N_\mathrm{corr}$ related to the activation energy of the relaxation time.

\begin{figure*}[htbp]
  \centering
  \includegraphics[width=\linewidth]{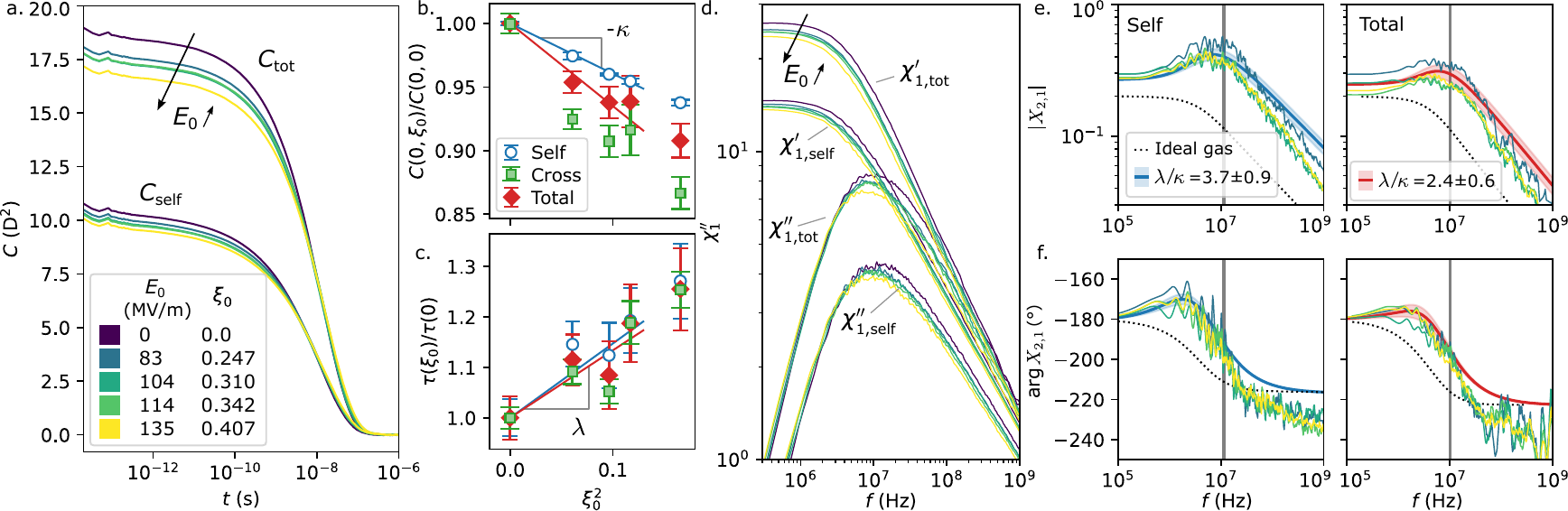}
 \caption{(a) Dipole correlation function for the self and total cases at $T = 263$~K, in the presence of a static electric field characterized by $\xi_0=\mu E_0/k_\mathrm{B}T$. (b) Static correlation and (c) relaxation time as a function of $\xi_0^2$. Solid lines are linear fits for $\xi_0<0.35$. (d) Real and imaginary parts of the self and total linear spectra $\chi_1$. (e) Modulus and (f) phase of the cubic susceptibility $X_{2,1}$ for the self part (left) and total (right). Thin lines were obtained directly from the linear spectra. Blue and red thick lines correspond to the $\mathrm{HN}_{\kappa,\lambda}$ method. The ideal gas of dipoles is shown with a dotted line. Vertical grey lines locate the peak frequency of the corresponding linear spectra.}
  \label{fig1}
\end{figure*}
The simulations were carried out in the NPT ensemble on a system of $N=2160$ glycerol molecules as in ref.~\cite{henotPCCP2023} with a force field previously employed in the literature~\cite{chelli1999glycerol, blieck_molecular_2005, egorov_molecular_2011, becher2021molecular}. A static electric field $E_0$ was applied along one direction of the simulation box with an amplitude characterized by a dimensionless coefficient $\xi_0 = \mu E_0/(k_\mathrm{B}T)$ where $k_\mathrm{B}$ is the Boltzmann constant and $\mu=3.2$~D is the mean molecule dipole. Care was taken to reach equilibrium at each temperature and field with an equilibration run of at least $100$ times the self dipole relaxation time $\tau_\mathrm{self}$ before a simulation run of at least $200\tau_\mathrm{self}$. The total dipole correlation function is defined as:
\begin{align}
    \label{eq_corr}
    C_\mathrm{tot}(t) = \frac{1}{N} \left\langle \sum_i\sum_j \vec{\mu}^\star_i(t_0) \cdot \vec{\mu}^\star_j(t_0+t) \right\rangle_{t_0}
\end{align}
Where $\vec{\mu}_i^\star = \vec{\mu}_i - \langle\vec{\mu} \rangle_{i,t}$ takes into account the polarization induced by the static field. $C_\mathrm{tot}$ can be separated into a self part $C_\mathrm{self}(t)$ for which $i=j$ and a cross part $C_\mathrm{cross}(t)$ corresponding to $i \neq j$~\footnote{Due to the effect of PBCs with the Ewald summation used in the treatment of Coulomb interactions~\cite{neumann1986computer, zhang2016computing}, $C_\mathrm{cross}$ is obtained by considering only pairs of dipoles closer than 7.5~\AA, see ref. ~\cite{henotPCCP2023}.}. 

The effect of $E_0$ on the self, cross, and total correlation functions is shown in fig.~\ref{fig1}a at $T = 263$~K which is the lowest temperature that we simulated. Increasing $E_0$ decreases the self, the cross and thus the total static correlation and increases the relaxation times $\tau_\mathrm{self}$, $\tau_\mathrm{cross}$ and $\tau_\mathrm{tot}$ (with $\tau = \int C(t)/C(0)dt$). For small enough field amplitude, it is expected that any effect should be proportional to $\xi_0^2$, which is the first non-zero order of the Taylor development for symmetry reasons: 
\begin{align}
    \frac{C(0, \xi_0)}{C(0, 0)} \approx 1-\kappa\xi_0^2 \qquad \frac{\tau(\xi_0)}{\tau(0)} \approx 1 +\lambda\xi_0^2
    \label{eq_kappa_lambda}
\end{align}
 These relations are tested in fig.~\ref{fig1}b and c, and appear valid within the uncertainty for $\xi_0 < 0.35$ ($E_0 < 114$~MV$\cdot$m$^{-1}$, only twice larger than highest experimental fields~\footnote{Experimentally in order to avoid dielectric breakdown, $E_0<70$~MV$\cdot$m$^{-1}$~\cite{lunkenheimer2017investigation}. Such fields are too weak to resolve non-linear effects in the simulation due to the limited system size and simulation time. At 263~K, the maximum field we used was 190~MV$\cdot$m$^{-1}$ for which small deviations from linearity start to be perceptible on $C(0)$. We thus determined $\kappa$ and $\lambda$ at this temperature by a fit on $E_0 < 114$~MV$\cdot$m$^{-1}$. At higher $T$, we kept $\xi_0<0.4$ (see fig.~\ref{appendix_fit_T})}).

The complex linear susceptibility $\chi_1 = \chi_1^\prime - j\chi_1^{\prime\prime}$ obtained from the Fourier transform of the correlation function (see appendix~A) is shown in fig.~\ref{fig1}d. The decrease of the static susceptibility $\chi_1^\prime(0)$ and of the peak frequency of $\chi_1^{\prime\prime}$ are well visible for the self and total cases.  For small enough $\xi_0$, the non-linear susceptibility $\chi_{2,1}$ can be approximated by the difference of the linear response in the presence and absence of field:
\begin{align}
\label{chi21_direct}
    \chi_{2,1}(f) \approx \frac{1}{3} \frac{\chi_{1}(f, E_0) -  \chi_{1}(f, 0)}{E_0^2}
\end{align}

The non-linear spectra can be expressed in a dimensionless form~\cite{lhote2014control, gadige2017unifying} $X_{2,1}(f) = \chi_{2,1}(f)k_\mathrm{B}T/(\chi_1(0)^2\epsilon_0v)$, where $v$ is the molecular volume, constructed to be temperature independent for non-interacting dipoles (see below). The modulus and phase of $X_{2,1}$ for the self and total cases are shown in fig.~\ref{fig1}e and f. They display the characteristic hump near $f_\alpha=1/(2\pi\tau_\mathrm{tot})$ observed experimentally in the deeply supercooled regime~\cite{lhote2014control, gadige2017unifying}.

To reduce the noise in the determination of $X_{2,1}(f)$, we use another approach, denoted in the following as $\mathrm{HN}_{\kappa,\lambda}$ and analogous to the one introduced in refs.~\cite{samanta2015dynamics, samanta2016electrorheological, young2016field, young2017nonlinear, gabriel2021high}. It relies on the assumption that $\chi_{2,1}$ results from two main effects: a decrease in static susceptibility ($\kappa>0$) and an increase in relaxation time ($\lambda>0$), with $\kappa$ and $\lambda$ fitted on the whole $E_0$ range where the effect is quadratic (see eq.~\ref{eq_kappa_lambda}). Additionally, the linear response is assumed to be described by a Havriliak–Negami expression (see Appendix B) whose parameters $\alpha$ and $\gamma$ are fitted on the linear spectra in the absence of field and assumed independent of it. In this framework, the presence of a hump on $|\chi_{2,1}(f)|$ and its amplitude are controlled by the ratio $\lambda/\kappa$. It exists only if $\lambda/\kappa$ is larger than a critical value close to $0.5$ (see fig.~\ref{appendix_HNmodel}). The resulting spectra are shown in fig.~\ref{fig1}e and f and compare very well to the one obtained from the direct method. The main difference concerns the high-frequency flank and is likely due to the assumption that $\gamma$ is independent of $E_0$. In the following, we use this second method, as it captures well hump amplitude, which is the focus of this study, and reduces the uncertainty on the peak value of the modulus.

\begin{figure}[htbp]
  \centering
  \includegraphics[width=\linewidth]{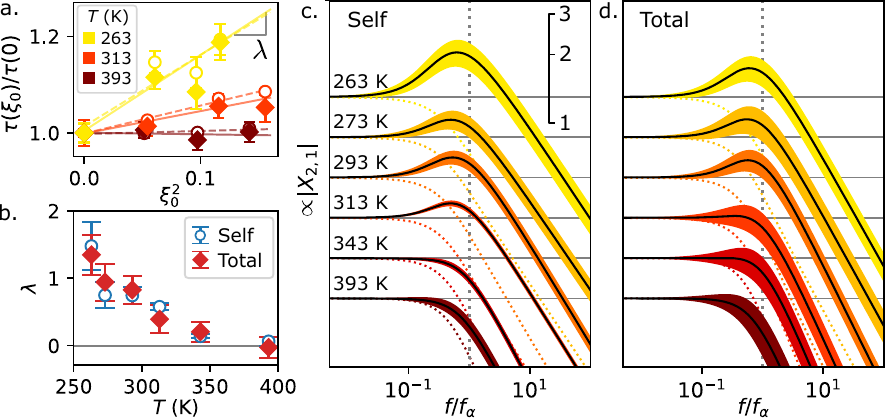}
 \caption{(a) Effect of $E_0$ on the relaxation time for three temperatures. Dashed and solid lines are linear fits for the self and total cases respectively with slopes $\lambda$ shown in (b) for all temperatures. (c-d) Modulus $|X_{2,1}|$ (from $\mathrm{HN}_{\kappa,\lambda}$) in log-log scale at different temperatures for the self (c) and total (d) cases. The uncertainty related to the determination of $\lambda/\kappa$ is shown. The curves are vertically shifted for readability.}
  \label{fig2}
\end{figure}

\begin{figure}[htbp]
  \centering
  \includegraphics[width=\linewidth]{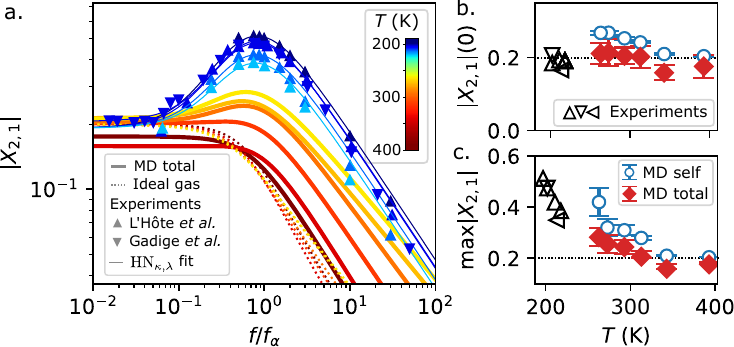}
 \caption{(a) Cubic modulus for the total MD case (thick lines) and from experiments~\cite{lhote2014control, gadige2017unifying} (markers). Thin lines correspond to a fit with $\mathrm{HN}_{\kappa,\lambda}$. (b) Static value and (c) maximum of the modulus for the MD self and total cases and from experiments~\cite{lhote2014control, samanta2015dynamics, gadige2017unifying} (see legend in fig.~\ref{fig4}a).}
  \label{fig3}
\end{figure}

We now turn to the effect of the temperature on the cubic spectra of glycerol, which we studied between 393~K and 263~K. The decrease in the static susceptibility with $E_0$, characterized by $\kappa$, appears to follow an inverse power-law temperature dependence (see fig.~\ref{appendix_kappa}) and is systematically higher for the total case than for the self by approximately 50~\%. In contrast, the effect of $E_0$ on the relaxation time is strongly affected by the temperature, as shown in fig.~\ref{fig2}a and b. It is smaller than the resolution at 393~K ($\lambda < 0.2$) and increases significantly upon cooling, reaching $\lambda=1.5$ at 263~K.
The modulus $|X_{2,1}|$ is displayed (up to a vertical prefactor for readability) in fig.~\ref{fig2}c and d for the self and total cases respectively. The feature standing out is the emergence and growth of a hump when the temperature is decreased. This effect can be quantified through the maximum of the modulus shown in fig.~\ref{fig3}c. The static value $|X_{2,1}(0)| = \mu^2\kappa/(3k_\mathrm{B}T\epsilon_0 v \chi_1(0))$ is shown in fig.~\ref{fig3}b and appears almost temperature independent, close to 0.2 for the total case while it slightly decreases with $T$ for the self case.

The present work is, to our knowledge, the first direct 3D MD simulation of the dielectric non-linear spectrum of a glass-forming liquid. For the present system, glycerol, the cubic response has been measured extensively~\cite{lhote2014control, samanta2015dynamics, gadige2017unifying}, and it is natural to directly compare these simulations with experimental observations. The moduli of $|X_{2,1}|$ obtained from refs.~\cite{lhote2014control, gadige2017unifying} are shown with markers in fig.~\ref{fig3}a. The static and maximum values deduced from experiments of refs.~\cite{lhote2014control, samanta2015dynamics, gadige2017unifying} (see appendix~D) are shown in fig.~\ref{fig3}b and c. In a dielectric experiment, the self and cross part cannot be separated and they should be compared to the MD total case. The range of accessible temperature in MD and experiments differs, the former being constrained by computational cost to $f>1$~MHz while the latter is limited to $f<50$~kHz. Still, the following points can be highlighted: (i) The static value $|X_{2,1}(0)|$ is independent of temperature and close to 0.2 in both total MD and experiments (see fig.~\ref{fig3}b). (ii) The MD and experimental hump grows consistently when the temperature is decreased. This is particularly visible in fig.~\ref{fig3}c where experimental and MD data align very well. (iii) In the $\mathrm{HN}_{\kappa,\lambda}$ method, the hump is controlled by the ratio $\lambda/\kappa$ and there again simulations and experiments align very well (see fig.~A4). As a consequence, combining MD and experiments allows to access, at least qualitatively, the evolution of the cubic spectra on a wide temperature range, from the liquid regime down to the glass transition. Some features are however not perfectly reproduced in the simulations. The amplitude of cross-correlations is underestimated which makes the linear dielectric response $\chi_1(0)$, and thus $\chi_{2,1}(0)$,  approximately 35-40~\% smaller than in experiments. The slope of the high-frequency flank of $|X_{2,1}(f)|$ is slightly smaller than in experiments as it is already the case on the linear spectra~\cite{henotPCCP2023}. This is not surprising as MD models usually have to be fitted specifically to reproduce dielectric quantities~\cite{zarzycki2020temperature} which is not the case here. Also, given the long time windows required, the model had to be kept fairly simple and electronic polarizability was not taken into account. Finally, the non-linear dielectric response of glycerol to a low frequency ($f<20$~kHz) ac field has recently been reported to display a strong increase for $T>280$~K~\cite{thoms2022strong}. This is in contrast with the almost temperature-independent $|X_{2,1}|(0)$ that we observed. This discrepancy may be related to the difference in non-linear observable (ac vs dc field) or to the fact that we are limited to frequencies not smaller than $10^{-2}f_\alpha$ while the increase was observed well below. 

Beyond these limitations, a strong point of this MD approach is that it gives access to both the self and cross parts. The interpretation of dielectric spectroscopy is complicated by the presence of the cross part whose dynamics differs from the self~\cite{gabriel2020intermolecular,koperwas2022computational, henotPCCP2023, alvarez2023debye, matyushov2023single}. Here, we can disentangle these effects for the cubic response and we can conclude that, for glycerol, the cross part has only a weak effect on the non-linear spectra. Indeed, the shift in relaxation time is identical for the the self, the cross and thus the total (see fig.~\ref{fig1}c and \ref{fig2}b) and the change in static susceptibility, while being affected by the cross-correlation (see fig.~\ref{fig1}b), keeps the same temperature dependence. Overall, as visible in fig.~\ref{fig1}e and \ref{fig3}c, the hump is only slightly more pronounced for the self than for the total. This confirms the pertinence of the cubic response in providing information on the molecular dynamics in itself for glycerol, independently of the existence of cross dipolar correlations. However, in some other systems, these cross-terms may dominate the cubic response~\cite{young2017nonlinear, gabriel2021high, sauer2024linear}. 

\begin{figure}[htbp]
  \centering
  \includegraphics[width=\linewidth]{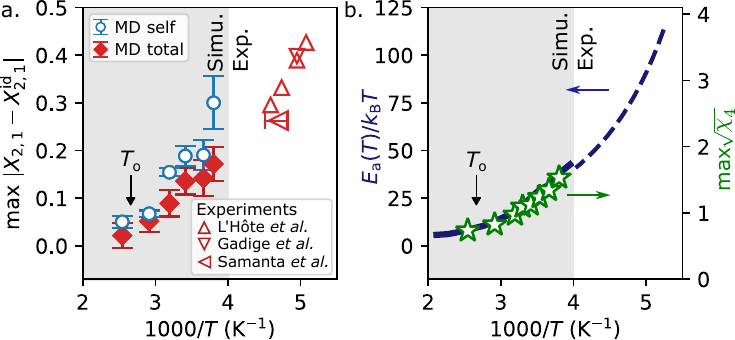}
 \caption{(a) Maximum of $|X_{2,1} - X_{2,1}^\mathrm{id}|$ (self in blue and total in red) as a function of the inverse temperature, from the present simulation and from experiments~\cite{lhote2014control, samanta2015dynamics, gadige2017unifying}. (b) Ratio of the apparent activation energy to the thermal energy (blue) from the simulation (solid line) and from experimental data~\cite{lunkenheimer2002dielectric} (dashed line). Right axis: maximum of $\sqrt{\chi_4}$ (green) describing the typical number of dynamically correlated molecules in the simulation.}
  \label{fig4}
\end{figure}

To investigate collective effects in the molecular dynamics, cubic spectra can be compared to the case of an ideal gas of dipoles (\textit{i.e.} without interactions)~\cite{dejardin2018nonlinear}. Consistently with the literature~\cite{casalini2015dynamic, albert2016fifth}, we computed the cubic response $\chi_{2,1}^\mathrm{id}$ of an ideal gas having the same linear response than the considered system (see appendix~E). These spectra are shown as dotted lines in fig.~\ref{fig1}e-f, \ref{fig2}c-d and \ref{fig3}a. Starting from $|X_{2,1}^\mathrm{id}(0)|=1/5$ very close to experimental and MD results, the modulus is strictly decreasing and systematically lower than $|X_{2,1}|$. The collective nature of the dynamics can be characterized through the maximum of $|X_{2,1}(f) - X_{2,1}^\mathrm{id}(f)|$~\cite{casalini2015dynamic, albert2016fifth}. This quantity has the advantage of being sensitive to small deviations from non-interacting dipoles, which is particularly interesting in the high-temperature regime accessed here. It is shown in fig.~\ref{fig4}a as a function of the inverse temperature for the present MD results (solid red diamonds) and for experiments (open red triangles)~\cite{lhote2014control, samanta2015dynamics, gadige2017unifying}. Here again, despite the difference in temperature range, simulation and experimental data seem to line up perfectly. 

A strength of these MD simulations is to give access to the high-temperature regime, not experimentally accessible. This enables us to observe the emergence of a hump in the cubic modulus between 343~K and 313~K. As mentioned above, a hump appears if $\lambda/\kappa$ exceeds a critical value. The parameter $\kappa$, characterizing the static response, does not strongly depend on temperature, and it is the field-induced slow-down of the dynamic $\lambda$, that, by appearing between 393 and 343~K, controls the emergence of the hump. Remarkably, it is in this same range, that the liquid goes from an Arrhenian dynamics with a constant activation energy $E_\infty$ to a super-activated behavior with an activation energy $E_\mathrm{a}(T)$ starting from    a plateau at $T>T_\mathrm{o}=375 \pm 10$~K, before increasing upon cooling~\cite{sastry2000onset, alba2002temperature} (see fig.~\ref{appendix_tau_lambda_kappa}). This onset temperature is higher than the fusion temperature of glycerol (292~K) suggesting that the progressive evolution of the free-energy landscape~\cite{berthier2011theoretical, baity2021revisiting}, eventually leading to the dynamical arrest at low temperature, starts well in the liquid regime~\cite{thoms2018breakdown}.

In MD simulations, dynamical correlations can be directly characterized though the maximum of the four-point susceptibility $\chi_4(t)$~\cite{berthier2011theoretical, casalini2015dynamic, pabst2024glassy}. This quantity characterizes spatial correlations between dynamical local correlation functions. As we focus on the cubic response of the orientational dynamics of molecular dipoles, we compute here $\chi_4(t)$ from the self dipole correlation function as in ref.~\cite{casalini2015dynamic} (see Appendix F). The link between $\chi_4$ and the number of dynamically correlated molecules $N_\mathrm{corr}$ was thoroughly studied by Berthier~\textit{et al.}~\cite{berthier2007spontaneous1, berthier2007spontaneous2} yielding $N_\mathrm{corr} \propto \sqrt{\mathrm{max} \chi_4(t)}$, which is shown in fig.~\ref{fig4}b (green stars) and can be superimposed with the cubic response. In experiments, $\chi_4$ is not accessible but a good proxy for $N_\mathrm{corr}$ is $\mathrm{d}\log \tau_\alpha/\mathrm{d}\log T = E_\mathrm{a}(T)/[k_\mathrm{B}T]$ (sometimes denoted $T\chi_T$)~\cite{dalle2007spatial, crauste2010evidence, brun2011nonlinear, lhote2014control, gadige2017unifying}. It is shown in fig.~\ref{fig4}b with blue lines for both the simulation and the experiments~\cite{lunkenheimer2002dielectric}. Both quantities can be superimposed, provided a small shift in vertical origin related to our choice of correlation function for $\chi_4$ (see Appendix F). It should be noted that some authors estimate $N_\mathrm{corr}$ from $E_\mathrm{a}(T)$~\cite{bauer2013cooperativity, pabst2024glassy} which yields similar results.

Overall, from fig.~\ref{fig4}a and b, it appears that all quantities show a similar behavior: the direct measure of the number of dynamically correlated molecules (from $\chi_4$), the change in activation energy $E_\mathrm{a}(T)/[k_\mathrm{B}T]$, and the evolution of the hump in the cubic response, either when considering only the self part or when including dipolar cross-correlations. For the later and the activation energy, this applies on a wide temperature range corresponding to 10 orders of magnitude in relaxation time. As mentioned in the introduction, using the cubic modulus as a probe of collective effect is debated in the literature~\cite{diezemann2012nonlinear, diezemann2013higher,  kim2016dynamics, diezemann2018nonlinear, speck2021modeling} and indeed there is no general theorem relating four-point correlation functions and cubic responses at finite frequencies. However, such a link has been discussed in the context of some microscopic theories~\cite{bouchaud2005nonlinear} and phenomenological models~\cite{ladieu2012toymodel, buchenau2017modeling} of the glass transition. In the context of the present work, we would like to emphasize one such approach related to the change of entropy induced by the static field $E_0$~\cite{johari2013effects, samanta2016electrorheological}. Combined with the Adam-Gibbs relation (linking $\tau$ and the configuration entropy $S_\mathrm{c}$), this leads to $\lambda \propto 1/S_\mathrm{c}^2$. As increasing $N_\mathrm{corr}$ corresponds to decreasing $S_\mathrm{c}$, it yields a growing cubic hump~\cite{gadige2017unifying} upon cooling.

In this letter, we have shown that the cubic spectrum of a realistic polar liquid, here glycerol, could be obtained from MD simulations. These results appear consistent with experimental measurements on the same system despite the difference in temperature range, which justifies the approximations made in the simulation. It turns out that the characteristic cubic hump, reported experimentally, emerges at the onset of super-activation and correlates well, at least qualitatively, with a direct characterization of the correlated dynamics in the simulation, and with a proxy of $N_\mathrm{corr}$ used in the literature~\cite{dalle2007spatial}. We evidenced that the cubic hump is controlled by the field induced slowing down of the dynamics, which itself may be related to the dynamical correlations. Besides, static cross-correlation effects can impact the shape of the cubic spectra although this remains a small effect in the case of glycerol.
As non-linear dielectric susceptibilities of supercooled liquid are considered a crucial tool to distinguish between different theoretical approaches of the glass transition~\cite{bouchaud2024dynamics, bertin2024nonlinear}, we hope that, in the future, this approach could be useful in the delicate interpretation of this powerful experimental technique.

\medskip

The authors are grateful to P.M. Déjardin, L. Berthier and C. Alba-Simionesco for fruitful discussions, to LABEX PALM, IRAMIS Institute and Paris-Saclay university for ﬁnancial support.

\bibliographystyle{apsrev4-2}
\bibliography{biblio}

\onecolumngrid

\vspace{12pt}
\noindent\hrulefill \hspace{24pt} {\bf End Matter} \hspace{24pt} \hrulefill
\vspace{12pt}

\twocolumngrid

\appendix
\medskip
\renewcommand\thefigure{A\arabic{figure}}    
\renewcommand{\theequation}{A\arabic{equation}}
\setcounter{equation}{0}
\setcounter{figure}{0}

\textit{Appendix A: Linear susceptibility.}
The complex linear susceptibility $\chi_1 = \chi_1^\prime - j\chi_1^{\prime\prime}$ was obtained through the following procedure: the imaginary part $\chi_1^{\prime \prime}(f) \propto f \times \mathrm{TF}(C(t))$ was computed from the Fourier transform $\mathrm{TF}$ of each correlation function using the fftlog algorithm adapted to log spaced data~\cite{Hamilton_2000}. From this, $\chi_1^\prime(f)$ was deduced using the Kramers-Kronig relations and the susceptibility was finally rescaled so that $\chi_1^\prime(0) = C(0)/(2\epsilon_0 k_\mathrm{B}T v)$. The present MD simulation uses Ewald summation to treat Coulomb interactions and the static permittivity $\epsilon(0) = 1 + \chi^\prime_1(0)$ is related to the Kirkwood correlation factor $g_\mathrm{K}$ (determined here not from the total simulation box with a cut-off radius such that the effect of electrostatic boundary conditions is not included) by~\cite{neumann1986computer, zhang2016computing}:
\begin{align}
    \frac{(\epsilon(0) - 1)(2\epsilon(0) + 1)}{\epsilon(0)} = \frac{\mu^2 g_\mathrm{K}}{\epsilon_0 k_\mathrm{B}Tv}
\end{align}
 The static correlation function can be expressed as $C_\mathrm{tot}(0) = \mu^2g_\mathrm{K}$. From this, $\chi^\prime_1(0)$ is given by the root of a second order polynomial and is dependence on $C_\mathrm{tot}(0)$ is not straightforward. In the case where $\chi^\prime_1(0) \gg 1$, $\chi^\prime_1(0)$ can be assumed proportional to $C_\mathrm{tot}(0)$: $\chi^\prime_1(0) \approx (\mu^2 g_\mathrm{K})/(2\epsilon_0 k_\mathrm{B}Tv)$
which is the expression used in the main text for the sake of simplicity. We have checked that the error introduced by this approximation is smaller than $3$~\%.

 \medskip

\begin{figure}[htbp]
  \centering
  \includegraphics[width=\linewidth]{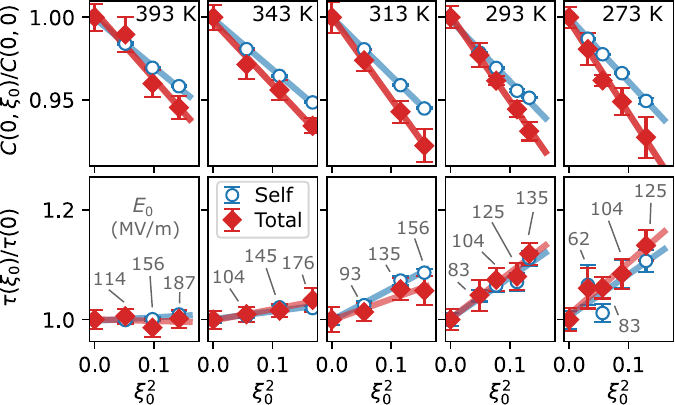}
 \caption{Evolution of the self and total static susceptibility and relaxation time with the electric field used to determine $\kappa$ and $\lambda$ for all the temperature studied, except 263~K already shown in fig.~\ref{fig1}. The values of the static electric field $E_0$, corresponding to $\xi_0\leq 0.4$, are given.}
  \label{appendix_fit_T}
\end{figure}

\begin{figure}[htbp]
  \centering
  \includegraphics[width=\linewidth]{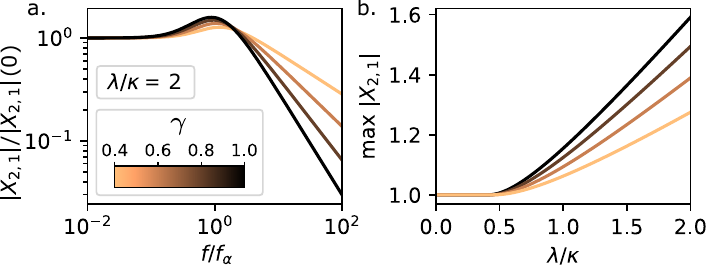}
 \caption{(a) Cubic modulus from $\mathrm{HN}_{\kappa,\lambda}$ for different high frequency slope controlled by the parameter $\gamma$. (b) Maximum of the modulus as a function of the ratio $\lambda/\kappa$.}
  \label{appendix_HNmodel}
\end{figure}

\textit{Appendix B: $\mathrm{HN}_{\kappa,\lambda}$ method.}  The expression used to parameterize the linear spectra is:
\begin{equation}
\label{eq_HN}
    \chi_{1, \mathrm{HN}}(f, \xi_0) = \frac{\chi_1(0)(1-\kappa\xi_0^2)}{(1+
    [j2\pi f \tau(1+\lambda\xi_0^2)]^\alpha)^\gamma}
\end{equation}
Within this framework, the emergence of a hump in the cubic modulus can be studied as a function of the parameters $\lambda$, $\kappa$, $\alpha$ and $\gamma$. As shown in fig.~\ref{appendix_HNmodel}, a hump is present if the ratio $\lambda/\kappa$ is larger than a critical value close to 0.5. The slope of the high frequency flank, controlled by $\gamma$, affects the growth of the hump with the ratio $\lambda/\kappa$ but has a negligible effect on the critical value.
 \medskip

\begin{figure}[htbp]
  \centering
  \includegraphics[width=\linewidth]{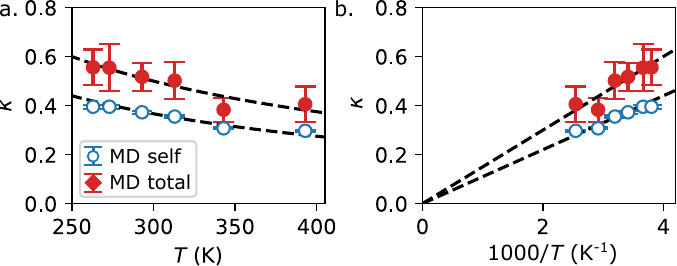}
 \caption{Static cubic response parameter $\kappa$, as a function of temperature (a) and inverse temperature (b), for the self and total cases. The black dashed lines correspond to $\kappa \propto 1/ T$.}
  \label{appendix_kappa}
\end{figure}

 \textit{Appendix C: Static cubic response.} The temperature dependence of the coefficient $\kappa$, characterizing the effect of the field on the static response, is shown in fig.~\ref{appendix_kappa} for the self and total case. It appears to follow an inverse temperature law (black dashed lines).

\medskip
\begin{figure}[htbp]
  \centering
  \includegraphics[width=\linewidth]{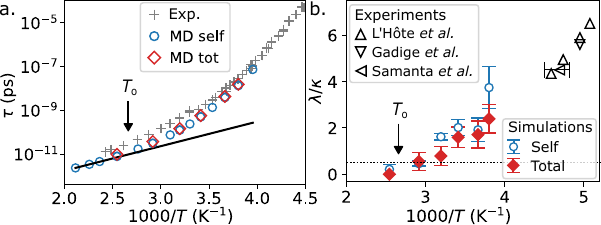}
 \caption{(a) Dipole relaxation time from MD simulations and from dielectric experiments~\cite{lunkenheimer2002dielectric}. The black line corresponds to a constant activation energy $E_\infty$. (b) Ratio $\lambda/\kappa$, as a function of the inverse temperature, for the MD self and total cases and from experiments~\cite{lhote2014control, samanta2015dynamics, gadige2017unifying}. The dashed horizontal line corresponds to the threshold leading to a hump (see fig.~\ref{appendix_HNmodel}).}
  \label{appendix_tau_lambda_kappa}
\end{figure}
\textit{Appendix D: Comparison with experimental data.} The modulus of $X_{2,1}$ from L'Hôte~\textit{et al.}~\cite{lhote2014control} and Gadige~\textit{et al.}~\cite{gadige2017unifying} were reproduced on fig.~\ref{fig3}a and the ratio $\lambda/\kappa$ was determined at each temperature by a fit with the $\mathrm{HN}_{\kappa,\lambda}$ expression. Samanta and Richert~\cite{samanta2015dynamics} measured the effect of a static field $E_0 = 22.5$~M$\cdot$V$^{-1}$ on the linear susceptibility $\chi_1(f)$. By fitting a Havriliak-Negami expression, they determined that between 208 and 227~K, the mean relaxation time was increased by $\Delta_{E_0}\ln\tau = 2.7$~\% and that the static susceptibility was reduced by $\Delta_{E_0}\ln \chi_1(0) = 0.6$~\%. The ratio of these two quantities is the same as the ratio $\lambda/\kappa = 4.5$. Besides, the static modulus can be computed from:
\begin{align}
    |X_{2,1}(0)| = \frac{1}{3}\frac{\Delta_{E_0}\ln\chi_1(0)}{E_0^2}\frac{k_\mathrm{B}T}{\chi_1(0)\epsilon_0 v}
\end{align}
The maximum of $|X_{2,1}(f)|$ and of $|X_{2,1}(f)-X_{2,1}^\mathrm{id}(f)|$ can be computed from $\lambda/\kappa$ and the HN parameters given by the authors.

 \medskip

\textit{Appendix E: Ideal gas of dipoles.} The cubic response to a static field of non-interacting dipoles for a Debye process is~\cite{dejardin2018nonlinear}:
\begin{align}
\label{eq_chi21_id_D}
    \chi_{2,1}^\mathrm{id,D}(x) = -\frac{1}{45}\frac{27+x^2-2x^4+jx(42+19x^2+x^4)}{(1+x^2)^2(9+x^2)}
\end{align}
with $x=2\pi f \tau$. The linear dynamics of the liquid can be decomposed on a basis of Debye functions and characterized by a distribution of relaxation time $G(\ln \tau)$ deduced, for example, from the HN fit~\cite{casalini2015dynamic}. The non-linear response $\chi_{2,1}^\mathrm{id}(f)$ of an ideal gas of dipole that would have the same linear response $\chi_1(f)$ is~\cite{casalini2015dynamic, albert2016fifth}:
\begin{align}
\label{eq_chi21_id}
    \chi_{2,1}^\mathrm{id}(f) = \int_{-\infty}^{\infty}{ \chi_{2,1}^\mathrm{id,D}(2\pi f\tau)G(\ln \tau})d\mathrm{ln}\tau
\end{align}
It should be noted that to obtain $\chi_{2,1}^\mathrm{id}$, the molecular reorientation was modeled here by rotational diffusion but other frameworks can lead to a ($T$ independent) humped shape for the modulus~\cite{diezemann2018nonlinear}. The fact that, at high temperature, $|X_{2,1}|$ does not display a hump and is very close to $|X_{2,1}^\mathrm{id}|$ justifies the simple choice made here.

\begin{figure}[htbp]
  \centering
  \includegraphics[width=\linewidth]{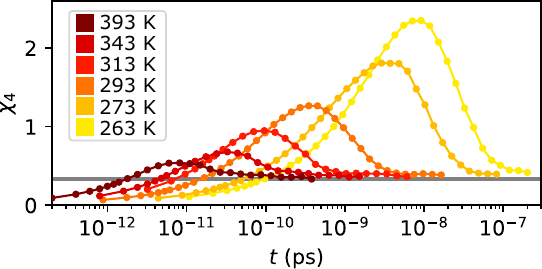}
 \caption{Four-point susceptibility computed from the dipole correlation. The grey horizontal line corresponds to the long-time expected value of 1/3.}
  \label{appendix_chi4}
\end{figure}

\medskip
\textit{Appendix F: Four-point susceptibility.} This quantity is defined by:
\begin{align}
\label{eq_def_chi4}
    \chi_4(t) = \frac{1}{N}\left\langle \sum_{j=1}^N c_i(t) c_j(t) \right\rangle_{i, t_0} - \langle c_i(t)\rangle^2_{i, t_0}
\end{align}
where $c(t)$ is a time correlation function chosen here to be the self dipole correlation: $c_i(t) = \vec{u}_i(t_0)\cdot \vec{u}_i(t_0+t)$ with $\vec{u}=\vec{\mu}/\mu$. The result for each temperature is shown in fig.~\ref{appendix_chi4}. It reaches its maximum close to the mean relaxation time and goes at long times to a plateau at $\approx 1/3$. This is related to the correlation function we choose, which, in the case $j=i$ in eq.~\ref{eq_def_chi4} gives, $\langle \cos^2\theta\rangle = 1/3$ for a random dipole orientation. This effect also manifest itself around $t=\tau_\alpha$ and explains the need for a vertical shift of the origin in fig.~\ref{fig4}b. 

\end{document}